# Open Challenges and Issues: Artificial Intelligence for Transactive Management

Asma Khatun*, Sk. Golam Sarowar Hossain

*Department of Computer Science & Engineering, Aliah University, Kolkata, India*

**ABSTRACT**

*The advancement of Artificial Intelligence (AI) has improved the automation of energy managements. In smart energy management or in a smart grid framework, all the devices and the distributed resources and renewable resources are embedded which leads to reduce cost. A smart energy management system, Transactive management (TM) is a concept to improve the efficiency and reliability of the power system. The aim of this article is to look for the current development of TM methods based on AI and Machine Learning (ML) technology. In AI paradigm, MultiAgent System (MAS) based method is an active research area and are still in evolution. Hence this article describes how MAS based method applied in TM. This paper also finds that MAS based method faces major difficulty to design or set up goal to various agents and describes how ML technique can contribute to that solution. A brief comparison analysis between MAS and ML techniques are also presented. At the end, this article summarizes the most relevant open challenges and issues on the AI based methods for transactive energy management.*

*Keywords: Multi Agent System; Transactive Energy; Smart Grid; Machine Learning; Distributed Energy; Literature Review*

## 1. INTRODUCTION

According to GridWise Architecture Council, transactive energy is ''a set of economic and control mechanisms that allow the dynamic balance of supply and demand across the entire electrical infrastructure using value as a key operational parameter'' (n. a., 2015). Transactive energy (TE) allows for faster transmission of information, including supply and demand quantities and prices, across the grid. It can manage both demand and supply sides which are designed for a changing environment with an increasing number of distributed resources and intelligent devices (Chen, 2017). Moreover, according to the NIST (n. a., 2019), the cost to supply electricity actually varies minutes by minutes.

TM has its own various disciplines such as smart grid (Romero, 2011, Reyasuddin, 2016), water pump (Sekar, 2015), telecommunications (Rahrah, 2015). Smart Grid, the way to future response, is different from the traditional grid since it facilitates two-way communications of electric data. Since smart grid can record data in real time with regards to electricity supply and real time demand during the transmission and distribution process of generation, transmission and distribution much more efficient and reliable, appealing to an optimized grid. The availability of two-way communication and the use of smart meters will enable online implementation and facilitate data collection within a fully automated system (Elkazz, 2016). A good comparison analysis of different components of traditional and smart grid has presented in the paper of (Romero, 2011). Artificial intelligent (AI) has the capacity to solve and take decision on real life scenarios to a known or unknown variant data. Hence deployment of Artificial intelligent (or machine learning) based techniques could be an alternative potential solution for an optimized grid management system. In the current literature, several AI based methods are proposed and are presented in the next section 2. The aim of this article is to present up-to-date survey of artificial intelligence based technique for transactive management. Rests of the paper is organized as follows. Section 2 presents the current state of the art technology on AI based methods providing a critical analysis of the relative merit, and potential pitfalls of the technique. Section 3 summarizes the article describing some of the open challenges as well as its future outlook.

*Corresponding author
E-mail addresses: asma.sun03@gmail.com

## 2. STATE OF THE ART METHODS

In the current development of the state-of-the-art technology many artificial intelligent based methods have been proposed. Those are included Genetic Algorithm (GA) (Elkazaz, 2016), Artificial Neural Networks (ANNs) (Chandler, 2014), Multi Agent System based method (MAS) (Reyasuddin, 2016, Stone, 2000). Interested researcher might find analytical survey papers (Romero, 2011, Nair, 2018) which reviews few AI based method in regards to the computational intelligence in the smart grid environment. Among all the aforementioned method stated, MAS based methods have achieved promising performance compared to the centralized energy management system on the state of the art technology.

In the application of smart grid energy management system, ANNs techniques are applied for various distributed resources for example, control management, solar energy heat-up response, solar irradiance, voltage stability monitoring or even for security issues. Among them ANNs techniques has been successfully applied in voltage stability monitoring (Venayagamoorthy, 2011) and control management (Chandler, 2014) application. In the method of (Chandler, 2014), they applied an integrated method of integer linear programming and ANN method. They manually implemented their method on real time SCADA (n.a., 2019) data. However, there need to more investigate on implementing and testing and with more data in an automated interface. In addition to that, the ANN techniques require a large amount of input data (Elkazaz,2016) proposed a method based on GA to apply in residence to reduce daily total operating cost by introducing different distributed grid (DG) units such as FC (fuel cell) and PV(rooftop photovoltaic) units at each home. As a comparing benchmark base scenario is used as a reference in which no FC and PV unit exist. There were total 4 scenario were performed in where each scenario 30 tests are performed. Given table 1 shows the comparison results of total daily cost of those scenarios.

*Table 1. Comparison results of total daily cost*

| Total Cost $C_{(TOTAL)}$ $ | Base | S1 | S2 | S3 | S4 |
|---|---|---|---|---|---|
| | 27.02 | 18.177 | 23.05 | 19.93 | 23.48 |

The above result indicates the difference in the total daily operating cost and the optimized result. However the proposed method has limitation to find the error as uncertainty associated with forecasted load has a great effect on the optimal results of the proposed system (Ren, 2008). On the other hand several agents perform several particular tasks in a MAS system. In MAS based methods, (Reyasuddin, 2016) used an optimization technique for the hybrid renewable energy system. In their method seven numbers of agents are used as actuators. As an example PV agent is used for measuring the total power output and similarly for other agents. However their report didn't cover for agent information in the case of learning process. They have shown higher performance and less power loss in the case of centralized energy management system.

There are broadly two classifications of multi agent structure namely centralized and decentralized. An extra dedicated coordinator agent exists in the centralized MAS. On the other hand agents communicate themselves peer to peer bases in decentralized MAS architecture (Weiss, 1999). A typical example of the difference between the above mentioned architecture has been shown in the following Figure 1 (a) & (b).

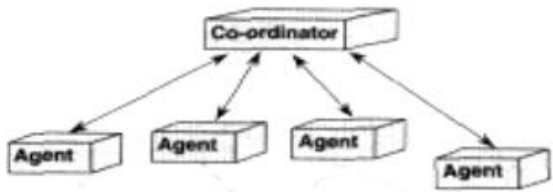 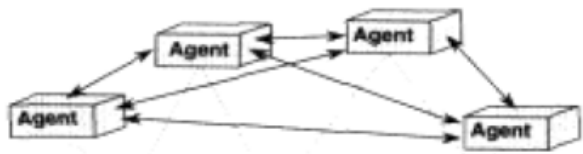

Fig. 1 (a) Centralized multi agent architecture        Fig. 1 (b) Decentralized multi agent architecture

In the current literature decentralized MAS is more popular than the former. In MAS based system it is important to consider the interaction among the multi agents. Otherwise in some cases the agent becomes greedy to act or run to achieve its goal (Stone, 2000). To solve this issue, Game theory technique has been applied by (Reyasuddin, 2016) for balancing the interaction among those multi agents. The game theoretic method is computationally complex as it is hard to find the equilibrium point (Dehghanpour, 2016). In MAS framework, an agent consists of several classes, layers and interfaces to communicate each other (Sardinha, 2004). It requires more attention for research to solve the communication and interaction among the agents. However, a priority queue based techniques can also be applied to become more robust. Although multi agent systems are highly adaptive to market and trading environment and easy to incorporate social abilities to exchange information, this method has several disadvantages. A fully multi agent-based system is much complex to modeling, maintaining and updating of the system or removal of any changes need to suffer.

Another challenge in multiple learning agents is to define a good learning goal. An agent must discover a solution on their own, using learning. Following are described how machine learning techniques are implemented in a distributed MAS system (Sardinha, 2004) for defining goal. The main aim of the method is to introduce cognition in agents through machine learning techniques that can leverage the performance of the system. Although integrating machine learning methods with MAS is a challenging task due to its inherent complexity in the interaction of multiple agents, this hybridization is not a new idea. From the past decades in many fields supervised machine learning techniques are introduced to MAS such as in semantic web (Williams, 2004), route planning (Gehrke, 2008) and BDI (Belief-Desire-Intentions) model (Airiau, 2008). The framework for introducing machine learning technique is described in the following figure 2. The learning problem is defined by two entities goal and performance measure. After defining the agent roles and the description of the agent behavior it is necessary to identify the best agents. Normally in a design of a Multi Agent System, each agent is concerned with a sub problem, which can be solved by applying a specific machine learning technique. The combination of these solutions must achieve the organization's goal and leverage the organization's performance measure. In the Agent Selection phase, the designer of the system notices that a specific agent has a complex plan to perform and needs a machine learning technique in order to improve the performance of the system. Moreover, for every selected agent, a problem domain analysis is performed to identify important learning issues. This phase is the Problem Domain Analysis and has the goal to establish a well-defined learning problem. More details can be found in (Sardinha, 2004).

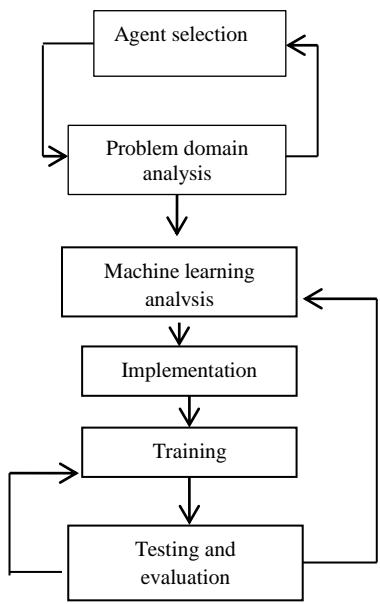

*Fig. 2. The methodology for introducing learning techniques* (Sardinha, 2004).

Other problem occurs as learning agents need to keep track of other learning agents which may create overlapping of each other coordination (Dehghanpour, 2016). To become more robust and sophisticated approach an example of approach has been seen which are integrated with machine learning and agent architecture (Khalil, 2015). In addition to that machine learning techniques are very autonomous for the decision making process and insensitive to market structure and large data sources (Chen, 2018). In the current development of the state-of-the-art methodology Reinforcement Learning (RL), a more powerful machine learning method is becoming hot research for transactive management energy system. In the approach of (Khalil, 2015) they implemented reinforcement machine learning technique in interactive multi agent systems namely MILIMAS in taxi domain and gained 80% improvement compared with traditional Q-learning (Watkins, 1992) algorithm for the same number of trials of the agents to reach to the passenger. Q-learning is a one kind of reinforcement learning which has no interaction of sharing information. Reinforcement learning allows generating a strategy for an agent in a situation, when the environment provides some feedback after the agent has acted. Feedback takes the form of a real number representing reward, which depends on the quality of the action executed by the agent in a given situation. The goal of the learning is to maximize estimated reward. The way of using sharing information in an interactive multi agent learning help to decide better and maximize the agents group achievement of goals. The framework for a normal agent which interact with environment and a framework for an agent model with reinforcement machine learning are presented in the Figures 3 and 4 respectively.

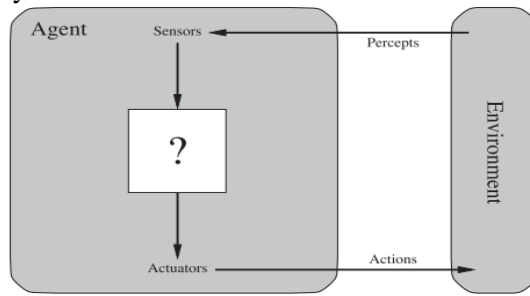

*Fig. 3. Agents interact with environments through sensors and actuators, adapted from (Russel, 2010)*

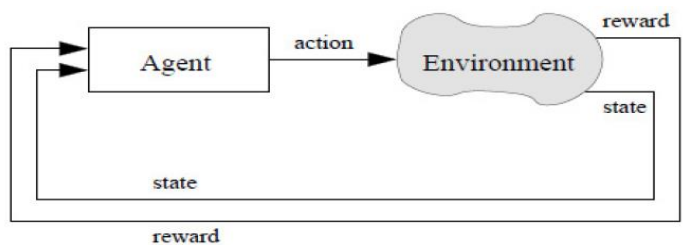

*Fig. 4. The reinforcement learning model in multi-agent systems, adapted from (Khalil, 2015)*

A modeling scheme with Markov's decision for transactive energy management in RL method has been seen in (Kim, 2018). It is noteworthy that, the RL method has the potential to deal with the energy trading problem and guide energy entities to interact with the market environment (Dehghanpour, 2016). The most important feature distinguishing RL from other types of learning is that it uses training information that evaluates the actions taken rather than instructions by giving correct actions (Sutton, 1998). In addition to that RL is more focused on goal directed learning from interaction (Sutton, 1998) which will also solve the challenging issue of multiple learning agents for defining a good learning goal. A summarized table of different modeling method is shown in Table 2 below.

*Table 2. A comparison summary of different method (Chen, 2018)*

| Method | Advantage | Disadvantage |
|---|---|---|
| MAS(Reyasuddin, 2016) | Parralellism makes speed up the system operation; Modulaity, adding new agent is easier; Furthermore identifying subtasks and assigning control in programming is simpler; parameters across the agents can be changed over time ; Lower in complexity | Most neglect transmission/distribution grid constraints; Results are mostly non-deterministic with poor interpretation; Not reliable due to external conditions and for policy makers; |
| Machine learning (Chandler, 2014) | Very autonomous decision-making process; Insensitive to market structure and large data sources; Medium complex | Data-driven and need realistic experiments; Usually need high computational resources; |
| GA(Elkazaz, 2016) | Simple to implement; performs better for highly constrained optimization problems, especially for problems having many local minima and non-smooth function; Low complexity | Uncertainty associated with forecasted load has a great effect on the optimal results; error need to be calculated; |

## 3. FUTURE WORKS AND SUMMARY

Transactive energy management gain popularity in recent studied in both academic and industry level. There exist very few articles on the state of the art artificial intelligence based method. Compare to other surveys, this paper covers the most recent and advanced work. In summary this article presented the current development of various artificial intelligence techniques and also look for the challenges occur in those methods. Comparative summary of the methods is also presented by describing merits and demerits. The way of integration of MAS and machine learning technique is also described briefly. As a future work, some of the proposed strategies are mentioned below:

- To adapt an unsupervised machine learning strategy such as (Stone, 2000) (as the data are unknown from previously) for transactive energy management. However, these learning techniques are not well suited for the multi agent as these are type of aimless learning (Khalil, 2015). But, a notable research found by developing hybrid learning approach which is computationally efficient and intelligent (Kiselev, 2008). The proposed approach is the application of multi agent algorithm for online agglomerative hierarchical clustering and is different from conventional unsupervised learning methods by being distributed, dynamic, and continuous. More details can be found in (Kiselev, 2008).

- To find a suitable evolution algorithm in order to make interaction among multi agents as MAS consists of several classes, layers and interfaces to communicate each other agents.

- To propose a hybrid technology with multi agent architecture and apply the clustering methods (machine learning techniques) with respect to learn new behavior in real time and to improve the system performance by facilitating the interaction dynamics among different agent entities (Elkazaz, 2016).

- To propose and establish a Reinforcement Learning based method for designing suitable goal as these methods is are most common algorithm for agent learning as they match the agent model exactly. In addition to that, Reinforcement learning allows an autonomous agent that has no prior knowledge of a task or an environment to learn its behavior by progressively improving its performance based on given rewards (Sadeghlou, 2014).